\documentclass[twoside,twocolumn,prb,a4paper,showpacs]{revtex4}

\usepackage{graphics}

\begin{document}

\title{Non-Markoffian effects of a simple nonlinear bath}

\author{Hanno Gassmann}
\email{Hanno.Gassmann@unibas.ch}
\author{Florian Marquardt}
\author{C. Bruder}
\affiliation{Department of Physics and Astronomy, University of Basel, Klingelbergstrasse 82, CH-4056 Basel, Switzerland}

\date{version of April 10th, 2002}

\begin{abstract}
We analyze a model of a nonlinear bath 
consisting of
a single two-level system coupled to a linear bath (a classical noise
force in the limit considered here). This allows us to study the
effects of a nonlinear, non-Markoffian bath in a particularly simple
situation. We analyze the effects of this bath onto the dynamics of a
spin by calculating the decay of the equilibrium correlator of the
spin's $z$-component. The exact results are compared with those
obtained using three commonly used approximations: a Markoffian master
equation for the spin dynamics, a weak-coupling approximation, and the
substitution of a linear bath for the original nonlinear bath. 
\end{abstract}


\maketitle

\section{Introduction}

The linear bath of oscillators plays a prominent role in discussions of dissipation
and decoherence \cite{1,2}. In the classical limit, the force fluctuations derived from
that bath correspond to a Gaussian random process. Although this is a generic
case (due to the central limit theorem), there are physical situations when
non-Gaussian random processes are important. In this report, we examine the
simplest possible quantum-mechanical bath whose fluctuations correspond to
a classical telegraph noise: a single two-level system subject to a white noise
force. The effects of this nonlinear bath are analyzed by coupling it, in turn,
to a spin, whose relaxational dynamics under the action of the bath is calculated.

\begin{figure}
{\par\centering \resizebox*{0,95\columnwidth}{!}{\includegraphics{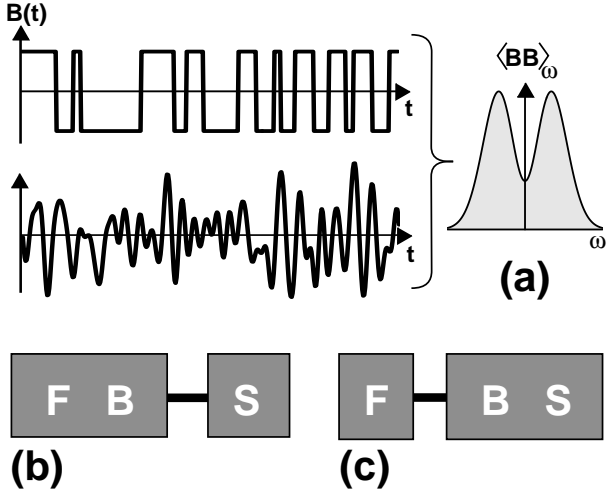}} \par}

\caption{\label{figur1} (a) Schematic representation of two stochastic processes corresponding to a classical two-level fluctuator,
or {}``telegraph noise{}'' (top), and a Gaussian process (bottom), yielding
the same power spectrum (right). (b) In our model, the two-level fluctuator
\protect\( B\protect \) is coupled to a noise force \protect\( F\protect \)
and therefore represents a (nonlinear and non-Markoffian) bath that acts on
a system \protect\( S\protect \). (c) The exact master equation description
(``approach 1'' in main text) treats \protect\( S\protect \) and \protect\( B\protect \)
as a composite system, subject to \protect\( F\protect \). }
\end{figure}

In the literature, another type of physically relevant nonlinear bath is usually
discussed: the spin bath\cite{stamp,sasha}, consisting of some large number of spins which are
coupled to the system under consideration. Our model system is simpler in that
it contains only a single {}``nonlinear element{}'', the two-level system.
Irreversibility is generated not by having a larger number of spins but by the
coupling to the linear bath. Although designed as a drastically simplified
model system, it may be physically relevant for situations like charged tunneling
systems\cite{paladino} in the vicinity of a mesoscopic quantum-coherent device (e.g. a Cooper-pair
box\cite{3}), which lead to electrostatic potential fluctuations and which are, themselves,
also subject to dissipation and decoherence by their environment. Viewed as
a whole, our model consists of two coupled two-level systems, one of which is coupled to a linear bath. Of course, such systems have been studied
before, both in the context of the quantum measurement problem and decoherence
of coupled qubit systems. In Refs.~\onlinecite{4},~\onlinecite{5},~\onlinecite{5b}
and ~\onlinecite{6}, the model of two spins (qubits) coupled to an environment has been analyzed in detail. However, we emphasize that our
perspective and the questions addressed in this article are different from these
approaches, since we are interested primarily in the differences arising from
substituting the nonlinear bath (in the form of the dissipative two-level system)
by a linear bath (see Fig. \ref{figur1}). This question is relevant, since, in many physical situations
where the precise nature of the bath decohering a given system is unknown, it
is simply treated as a linear bath, with some given correlation function. It
is, therefore, desirable to understand in more detail the kind and magnitude
of possible errors introduced by such an approximation, in cases where the coupling cannot be assumed to be very weak.

The basic strategy is to calculate the equilibrium correlator of the
two-level system exactly (which can be done in the limit
of infinite temperature) and to compare the results to three common 
approximations. One of those involves replacing the nonlinear bath 
by a linear bath, whose correlation function is prescribed to be the
same as that of the nonlinear bath. The others are a Markoffian master
equation and a weak-coupling approximation, applied to the dynamics of
the spin under the influence of the bath.

The remainder of this work is organized as follows: In Sec. \ref{model}, we
give the model Hamiltonian. In Sec. \ref{master}, the four different approaches
are defined by specifying the evolution equation for the density
matrix in each case. Afterwards, we explain (Sec. \ref{relax}) how the
equilibrium correlator of the \(z\)-component of the spin may be obtained by
solving these equations. Finally, Sec. \ref{num}. presents plots showing the
numerical results for the spin correlator, along with a comparison
between the different approaches.

\section{The model}\label{model}

We consider a two-level system \( S \) coupled to another two-level system
\( B \), which represents an example of a nonlinear dissipative bath, since it is
subject to a fluctuating force \( F \): 

\begin{equation}
\hat{H}=\epsilon _{S}\hat{\sigma }^{S}_{z}+\Delta _{S}\hat{\sigma }^{S}_{x}+J\hat{\sigma }^{S}_{z}\hat{\sigma }^{B}_{z}+\Delta \hat{\sigma }^{B}_{x}+\hat{F}\hat{\sigma }_{z}^{B}+\hat{H}_{F}.
\end{equation}

Here the parameters \( \epsilon _{S} \) and \( \Delta _{S} \) serve to define
any desired two-level system \( S \). This system is coupled to \( B \) via
\( \hat{\sigma }^{B}_{z} \), with the coupling strength between \( S \) and
\( B \) being given by \( J \). The oscillations of \( \hat{\sigma }^{B}_{z}(t) \)
at the frequency \( 2\Delta  \) are noisy, due to the action of the fluctuating
force \( \hat{F} \), which may stem from a linear bath of harmonic oscillators,
whose Hamiltonian is given by \( \hat{H}_{F} \). Below we will specialize to
the limiting case of infinite temperature, where \( \hat{F} \) becomes a purely
classical noise force.

The dissipative dynamics of \( S \) can be characterized in terms of several
different quantities. Here we will analyze  the decay of the
equilibrium correlator \( \left\langle \hat{\sigma }_{z}^{S}(t)\hat{\sigma }_{z}^{S}(0)\right\rangle  \).

Solving the full model of a system of two interacting spins coupled to a linear
bath at arbitrary temperatures and coupling strengths represents a formidable
problem in itself. It has been attacked in the past using the Feynman-Vernon
influence functional \cite{7}, both analytically \cite{4} (in certain limiting cases)
as well as numerically \cite{6}. For
our purposes, we will be content with analyzing a technically simpler special case. We choose
\( F \) to be a classical white noise fluctuating force, 

\begin{equation}
\left\langle F(t)F(0) \right\rangle = \gamma \delta(t)\, \label{FF},
\end{equation}

which corresponds
to the limit of infinite temperature \( T \) of the bath (taken such that the overall noise strength remains constant).

Under these circumstances, the dissipative dynamics of \( S+B \) under
the action of \(F\) can be described exactly by
using a Markoffian master equation. Note that this is, of course, unrelated to the validity
of a master equation description for the action of \( B \) onto \( S
\) alone, which we will discuss below.
The limit of infinite temperature is dictated mostly by the desire to have a
comparatively strong decay of the correlator of \( \hat{\sigma }^{B}_{z}(t) \)
(with a decay rate on the order of B's transition frequency \( 2\Delta  \))
while still retaining the validity of a simple Markoffian master equation description (for the
full system \( S+B \)).

In the following, we will call the exact solution ``approach~1'',
while ``approach~2'' refers to a simple master equation applied to $S$
alone, ``approach~3'' replaces the nonlinear by a linear bath, and
``approach~4'' is the weak-coupling approximation.

\section{Master equation description of the relaxation}\label{master}

\emph{Master equation for approach 1:} 
The following exact master equation description is used for the action of  \( F \) onto the combined system \( B+S \):

\begin{equation}
\label{mastereqnonsec}
\dot{\hat{\rho}}_{SB}(t)=-i[\hat{H}_{SB},\hat{\rho}_{SB}(t)]-\gamma\hat{\rho}_{SB}(t)+\gamma\hat{\sigma }^{B}_{z}\hat{\rho}_{SB}(t)\hat{\sigma }^{B}_{z}.
\end{equation}
\(\hat{H}_{SB}\) is the Hamiltonian for the system \(S\) and \(B\)
alone.
Note that, in contrast to the usual master equation, {\em no} secular
approximation\cite{8} has been used in deriving this equation, which means the
resulting decay rate does not have to be small when compared to the
transition frequencies of the system $S+B$. This is possible because
the bath correlation function is a delta function, which also makes the equation exact. We remark further that
Eq. (\ref{mastereqnonsec}) is solved directly in the basis 
in which $\hat{\sigma}_z^B$ and $\hat{\sigma}_z^S$ are diagonal. A transformation
into the interaction picture (as is commonly performed for the usual 
master equation description) would lead to explicitly time-dependent
terms in this equation. 

\emph{Master equation for approach 2:} 
 As has been explained above, we will use the master equation description
not only for the action of \( F \) onto the combined system \( S+B \)
(see  Eq. (\ref{mastereqnonsec})), but also for the action of \( F+B
 \) onto \( S \) alone. This constitutes the approximate
 ``approach~2'', involving the usual kind of master equation, which is valid only for
sufficiently weak coupling J, since it is derived by applying both the
Markoff and secular approximation (see Ref.~\onlinecite{8}). In the unperturbed 
eigenbasis of system $S$, it reads:

\begin{eqnarray}
\label{mastereq}
&&\dot{\rho }_{S kj}=-(\Gamma _{k}+\Gamma _{j}+\tilde{\Gamma
}_{kj}+i(\Delta _{k}-\Delta _{j})
\\
&&+i(E_{k}-E_{j}))\rho _{S kj}+\delta _{kj}\sum _{l\neq k}\rho _{S ll}|A_{kl}|^{2}2\pi \left\langle BB\right\rangle _{E_{l}-E_{k}}.\nonumber
\end{eqnarray}

Equation (\ref{mastereq}) describes the relaxation of the reduced density
matrix $\hat{\rho}_{S}$ of system \(S\) alone, under the action of the
coupling $J\hat{\sigma}^S_z\hat{\sigma}^B_z$ to the bath. We have
introduced the abbreviation $\hat A\equiv\hat{\sigma}^S_z$.

 The Fourier transform of the correlator of $\hat{B}\equiv J\hat{\sigma}^B_z$ defines the
``bath spectrum'':
\begin{eqnarray}
\left\langle
B(t)B(0)\right\rangle\equiv\left\langle\hat{B}(t)\hat{B}(0)\right\rangle=J^2\left\langle
\hat{\sigma}^B_z(t)\hat{\sigma}^B_z(0)\right\rangle\nonumber\\
\left\langle BB\right\rangle _{\omega }\equiv \frac{1}{2\pi }\int _{-\infty }^{+\infty } dt\, e^{i\omega t }\left\langle
B(t)B(0)\right\rangle \, \label{BBB}.
\end{eqnarray}

It is real and symmetric in the limit of infinite temperature
considered here, and therefore it is equivalent to a classical colored noise force. As will
be explained below, $\left\langle BB\right\rangle _{\omega }$ is found
by applying the master equation (\ref{mastereqnonsec}) to $B$ alone.
The decay rates are defined by

\begin{eqnarray}
\Gamma _{k} & \equiv  & \pi \sum _{n}|A_{kn}|^{2}\left\langle BB\right\rangle _{E_{k}-E_{n}}\nonumber \\
\tilde{\Gamma }_{kj} & \equiv  & -2\pi A_{kk}A_{jj}\left\langle BB\right\rangle _{0}\, ,
\end{eqnarray}

and the energy shifts are given via

\begin{equation}
\Delta _{k}\equiv \sum _{n}|A_{kn}|^{2}\int d\omega \frac{\left\langle BB\right\rangle _{\omega }}{E_{k}-E_{n}-\omega }.
\end{equation}

Here the indices and energies refer to the unperturbed eigenstates of the original Hamiltonian of $S$ alone: \( \hat{H}_{S}\equiv \epsilon _{S}\hat{\sigma }^{S}_{z}+\Delta _{S}\hat{\sigma }^{S}_{x} \).

{\em Approach 3} consists in replacing the nonlinear bath by a linear
one. If the two-level
fluctuator \(B\) were replaced by a harmonic oscillator\cite{garg,fkw}, this procedure
of substituting a linear bath with an appropriate correlation
function for the combination of $F$ and $B$ would be exact. Here, it
is
an approximation whose reliability we want to analyze by comparison to
the exact solution.
In our case, the fact that the power spectrum $\left\langle BB\right\rangle _{\omega }$ is real and symmetric means that $B$ can be treated as a classical Gaussian random 
process. Therefore, we have to solve
\begin{equation}
\dot{\hat{\rho}}_{S}(t)=-i[\hat{H}_{stoch}(t),\hat{\rho}_{S}(t)]
\end{equation}
with the stochastic time-dependent Hamiltonian: 
\begin{equation}\label{htime}
\hat{H}_{stoch}(t)=\epsilon _{S}\hat{\sigma }^{S}_{z}+\Delta _{S}\hat{\sigma }^{S}_{x}+B(t)\hat{\sigma }^{S}_{z}.
\end{equation}
The numerical procedure used for averaging over the realizations of the
process \(B(\cdot)\) is described below, in Sec. \ref{num}.

{\em Approach 4:}
Instead of the Markoff approximation one can use a weak-coupling
approximation. This keeps the full information contained in the
correlator \(\left\langle BB\right\rangle _{\omega }\), at the price of
introducing a kernel for the master equation which is no longer local
in time.
We find the following weak-coupling equation, where only terms up to
second order in \(J\) have been kept:
\begin{eqnarray}
\label{weak}
&&\dot{\hat{\rho}}_{S}(t)=-i[\hat{H_{S}},\hat{\rho}_{S}(t)]\\
&&-\int_{0}^{t}
d\tau\left\langle
B(\tau)B(0) \right\rangle
\Big\lbrack
\hat{\sigma}^{S}_{z},e^{-i\hat{H_{S}}\tau}
\Big\lbrack
\hat{\sigma}^{S}_{z},\hat{\rho}_{S}(t-\tau)
\Big\rbrack
e^{i\hat{H_{S}}\tau}
\Big\rbrack.\nonumber
\end{eqnarray}
This equation is conveniently
solved by using the Laplace transform.
\section{Relaxation of density matrix and correlator}\label{relax}

\emph{Decay of the equilibrium correlator} - We want to obtain the equilibrium
correlator of \( \hat{\sigma }_{z}^{S}(t) \),

\begin{equation}
\label{corrS}
\left\langle \hat{\sigma }^{S}_{z}(t)\hat{\sigma }^{S}_{z}(0)\right\rangle \equiv tr\left[ \hat{\rho }^{(eq)}_{SB}\hat{\sigma }^{S}_{z}(t)\hat{\sigma }^{S}_{z}(0)\right] \, .
\end{equation}

It is convenient to rewrite Eq. (\ref{corrS}) in terms of the projector
onto the spin-up state of \( S \), \( \hat{P}\equiv \left| \uparrow
\right\rangle _{S}\left\langle \uparrow \right| _{S} = \frac{1}{2}(1+\hat{\sigma }^{S}_{z})\): 

\begin{equation}
\left\langle \hat{\sigma }^{S}_{z}(t)\hat{\sigma }^{S}_{z}(0)\right\rangle =4\left\langle \hat{P}(t)\hat{P}(0)\right\rangle -1.
\end{equation}

Here, we have used \(\hat{\rho }^{(eq)}_{SB}=\frac{1}{4}\).
The correlator of \( \hat{P}(t) \) can be found by calculating the
probability to find the system \(S\) in the state ``up'' at the time
\(t\), if it had been ``up'' at time \(0\). This has to be averaged
over all realizations of the random process \(F(\cdot)\):

\begin{equation}
\label{PP}
\left\langle \hat{P}(t)\hat{P}(0)\right\rangle =\frac{1}{2}tr_{B}\left\langle \left\langle \uparrow \right| _{S}\hat{U}_{F}(t)\, \hat{P}\otimes \hat{\rho} ^{(eq)}_{B}\, \hat{U}_{F}^{\dagger }(t)\left| \uparrow \right\rangle _{S}\right\rangle _{F}.
\end{equation}

Here \( \hat{U}_{F}(t) \) is the time-evolution operator for \( S+B \) under
the action of a given realization of \( F(\cdot ) \). This equation is valid only because, in our model, the probability of finding ``spin up'' at a certain instant of time is independent of the history of \( F(\cdot ) \). The expression (\ref{PP})
is nothing but the population \( \rho _{S11}(t) \) of the state \( \left| \uparrow \right\rangle _{S} \)
for a time-evolution starting from the initial condition of {}``spin up{}'',
\( \hat{\rho }_{SB}(0)=\hat{P}\otimes \hat{\rho }^{(eq)}_{B} \): 

\begin{equation}
\label{PPEQ}
\left\langle \hat{P}(t)\hat{P}(0)\right\rangle =\frac{1}{2}\rho _{S11}(|t|).
\end{equation}

Note that \( \rho _{S11} \) decays towards \( 1/2 \), such that \( \left\langle \hat{\sigma }^{S}_{z}(t)\hat{\sigma }^{S}_{z}(0)\right\rangle  \)
vanishes for \( t\rightarrow \infty  \) (as it should be). We have used the fact that the correlator is symmetric in time, since the potentially 
antisymmetric imaginary part vanishes. \( \hat{\rho }_{S}(t) \)
can be calculated by applying the master equation that describes the action
of \( F \) onto \( S+B \). Put differently, Eq. (\ref{PPEQ})
constitutes an example of the quantum regression theorem. Using \(
\hat{\rho }_{S} \), we calculate the Fourier transform of the equilibrium correlator of
\( \hat{\sigma }_{z}^{S}(t) \):

\begin{eqnarray}
\label{Kzz}
K_{zz}^{S}(\omega ) & \equiv & \frac{1}{2\pi }\int _{-\infty }^{+\infty }dt\, e^{i\omega t}\left\langle \hat{\sigma }_{z}^{S}(t)\hat{\sigma }_{z}^{S}(0)\right\rangle \nonumber \\
&=&\frac{1}{\pi}\int_{-\infty}^{+\infty} dt\, e^{i\omega t} (\rho_{S11}(|t|)-1/2) .
\end{eqnarray}

\( K^{S}_{zz}(\omega ) \) is real-valued, symmetric and the integral over all
frequencies gives \( 1 \).

The master equation for the density matrix \( \hat{\rho } \) (\( \equiv \hat{\rho }_{SB} \))
in the four-dimensional Hilbert space of \( S+B \) represents a system of linear
differential equations with constant coefficients. The latter are given by a
complex-valued \( 16\times 16 \) matrix \( C \) that corresponds to the {}``superoperator{}''
on the right-hand side of the master equation. The solution is the complex vector
\( \rho  \), which consists of the \( 16 \) components of the density matrix
\( \hat{\rho } \) :

\begin{equation}
\label{mastersuper}
\dot{\rho }=-C\rho \, .
\end{equation}

The entries of \( C \) can be read off directly from Eq. (\ref{mastereqnonsec}).
The formal solution of Eq. (\ref{mastersuper}),

\begin{equation}
\rho (t)=e^{-Ct}\rho (0)\, ,
\end{equation}
 can be expressed in terms of the right-eigenvectors \( \left| \rho ^{(j)}\right\rangle  \), the left-eigenvectors  \( \left\langle \rho ^{(j)}\right|\)
and the eigenvalues \( \lambda ^{(j)} \) of \( C \):

\begin{equation}
\rho (t)=\sum _{j}\left| \rho ^{(j)}\right\rangle \left\langle \rho ^{(j)}| \rho (0)\right\rangle e^{-\lambda ^{(j)}t}.
\end{equation}

\( C \) is not necessarily hermitian, so that the \( \lambda ^{(j)} \) usually
are complex-valued (with non-negative real parts) and the \( \left| \rho^{(j)}\right\rangle  \)
do not form an orthonormal basis (however, \( \left\langle \rho ^{(i)} \right| \left. \rho ^{(j)}\right\rangle =\delta _{ij} \)
by construction). In order to obtain \( \rho _{S11}(t) \), we have to perform
the trace over \( B \), \( \rho _{S11}(t)=\rho _{SB1111}(t)+\rho _{SB1212}(t) \).
(In \( \rho _{SBs'b'sb} \) the indices \( s,s' \) refer to \( S \), while
\( b,b' \) refer to \( B \).) We will use the same notation for the components
of \( \rho ^{(j)} \), which is a complex vector. Then we obtain 

\begin{equation}
\label{Requ}
\int _{0}^{\infty }dt\, e^{i\omega t}\rho _{S11}(t)=\sum _{j}(\rho ^{(j)}_{1111}+\rho ^{(j)}_{1212}){\left\langle \rho ^{(j)} | \rho (0)\right\rangle \over \lambda ^{(j)}-i\omega }.
\end{equation}

Taking the real part of this expression gives \(K_{zz}^{S}(\omega )\), see
Eq. (\ref{Kzz}).

An analogous formula holds for the master equation describing the decay of \( \hat{\rho }_{S} \) (\( \equiv \hat{\rho}  \)
in that case)
under the action of \( F+B \) (approach 2), see
Eq. (\ref{mastereq}). Then, \( C \) corresponds to the \( 4\times 4 \)
matrix whose entries are read off from Eq. (\ref{mastereq}).
Therefore, Eq. (\ref{Requ}) only contains \( \rho ^{(j)}_{11} \) in that case,
instead of the sum inside the brackets.

Similarly, we have to obtain the equilibrium correlator of \( \hat{\sigma }^{B}_{z}(t) \),
which is needed as input both for the master equation describing the relaxation of
\( S \) alone (approach~2), the numerical sampling of random
processes (approach~3) and the weak-coupling approximation (approach~4).

\begin{figure}
{\par\centering \resizebox*{0,95\columnwidth}{!}{\includegraphics{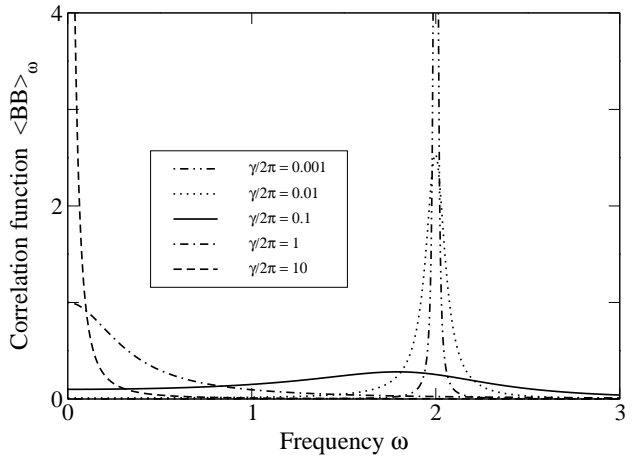}} \par}
\caption{\label{fig:bbcorr}The \(\left\langle BB \right\rangle_{\omega
}\) correlation function for different values of \(\gamma\). The
parameters are: \(J=1, \Delta=1
\) }
\end{figure}

This is done by calculating the relaxation of \( \hat{\rho }_{B}(t) \) under
the action of \( F \), starting from the initial condition \( \hat{\rho }_{B}(0)=\left| \uparrow\right\rangle_B \left\langle \uparrow \right|_B \)
and applying the same formulas as above (with \( B \) instead of \( S \)),
for the master equation (\ref{mastereqnonsec}), adapted to the two-dimensional Hilbert space of \( B \) (with
a \( 4\times 4 \) matrix \( C \)). \( \left\langle \hat{\sigma }_{z}^{B}(t)\hat{\sigma }^{B}_{z}(0)\right\rangle  \)
undergoes damped oscillations. Its Fourier transform 

\begin{equation}
\label{BB}
 K^{B}_{zz}(\omega )=\frac{\left\langle BB \right\rangle_{\omega }}{J^2}= \frac{8 \Delta ^2 \gamma}{\pi}\frac{1}{(\omega ^2-4 \Delta ^2 )^2 + 4 \omega ^2 \gamma ^2}
\end{equation}

consists of broad
peaks of width $\gamma$ (for \(\gamma^2<4\Delta^2\)), which is proportional to the strength of the
noise force $F$ and may be comparable to the transition frequency \(
2\Delta  \) itself (see Fig.~\ref{fig:bbcorr}).
Thus, \( B \) indeed represents a noisy two-level fluctuator, which acts onto
\( S \) as a \emph{nonlinear} (non-Gaussian) and \emph{non-Markoffian}
(colored) bath.

In order to solve the equations (\ref{weak}) of approach 4
(weak-coupling approximation), we need the Laplace transform of the equilibrium correlator of the
bath \(B\), which is connected to the Fourier transform in the usual way:
\begin{eqnarray}
C_{BB}(s)&\equiv&\int_{0}^{\infty}dt e^{-s t}\left\langle
B(t)B(0)
\right\rangle\\&=&\int_{-\infty}^{\infty}d\omega\frac{\left\langle B B
\right\rangle_{\omega}}{s+i\omega}=J^2\frac{s+2\gamma}{s^2+2\gamma s+4\Delta^2}.\nonumber
\end{eqnarray}
Using the Laplace transform, the system of differential equations becomes a system of
linear algebraic equations, which can be solved by matrix inversion. 
All the results can be obtained analytically. However, here we only present
the comparatively brief expression  for the special case of
\(\epsilon_{S}=0\):
\begin{equation}\label{lapsol}
 K^{S}_{zz}(\omega )=\frac{1}{\pi}\Re e\Big\{\frac{s+4
 C_{BB}(s)}{s^2+4 C_{BB}(s)+4\Delta_{S}^2}|_{s=-i\omega}\Big\}.
\end{equation}

\section{Numerical results}\label{num}

The following steps have been performed for calculating the correlator $K_{zz}^S(\omega)$ of \( \hat{\sigma }^{S}_{z}(t) \):

\emph{Approach 1 (``exact description''):} The entries of the matrix
$C$ are obtained from Eq. (\ref{mastereqnonsec}). The eigenvalues and
eigenvectors of \( C \) are calculated numerically and used to get $K_{zz}^S(\omega)$
according to Eqs. (\ref{Requ}) and (\ref{Kzz}).

\emph{Approach 2 (``simple master equation for S''):} First the action
of \( F \) onto \( B \) is considered, to obtain the correlation
function \( \left\langle BB \right\rangle_{\omega } \). This result is
given in Eq. (\ref{BB}). It is used to set up the master equation
describing the action of $F+B$ onto $S$, Eq. (\ref{mastereq}). Its
coefficients define a $4\times4$ ``C-matrix'', which is
diagonalized. The results are inserted into the appropriately modified
Eq. (\ref{Requ}), in order to obtain $K_{zz}^S(\omega)$.

 \emph{Approach 3 (``numerical sampling using colored noise''):} We
 calculate numerically the time-evolution of $\hat{\rho}_S(t)$ under
 the action of the stochastic time-dependent Hamiltonian defined in
 Eq. (\ref{htime}), which depends on \(B(t)\).

\( B(t) \) is a particular realization of a stationary Gaussian random process of zero mean and power spectrum $\left\langle BB \right\rangle_{\omega }$. The density matrix $\hat{\rho}_S(t)$ has to be averaged over a statistical sample of  different field configurations \(B(t)\). This sample is produced by generating the Fourier coefficients of $B$ as independent complex Gaussian random variables of appropriate variance (given by the power spectrum). The field $B(t)$ itself is obtained using a Fast Fourier Transform (FFT). After averaging, we may use

\begin{equation}
\left\langle \hat{P}(t) \hat{P}(0) \right\rangle={1 \over 2} \left\langle \rho_{S11}(|t|) \right\rangle_{B} \, \label{PPP}
\end{equation}

and Eqs. (\ref{PP}) and (\ref{Kzz}) in order to obtain $K_{zz}^S(\omega)$. To this end, the Fourier transform of $\rho_{S11}(|t|)$ is calculated numerically, using a FFT on a time-grid of sufficiently small step-size $\Delta t$ and sufficiently large length. The results displayed in the figures have been obtained using $10^4$ samples and a frequency resolution of $\Delta\omega=2\pi / 800$. The curves have been smoothed by averaging over $5$ to $20$ adjacent frequency bins. 

 \emph{Approach 4 (``weak-coupling equation''):} We evaluate the
 analytical results obtained above (see, e.g., Eq. (\ref{lapsol})) with the
 appropriate numerical values.

\begin{figure}

{\par\centering \resizebox*{0,95\columnwidth}{!}{\includegraphics{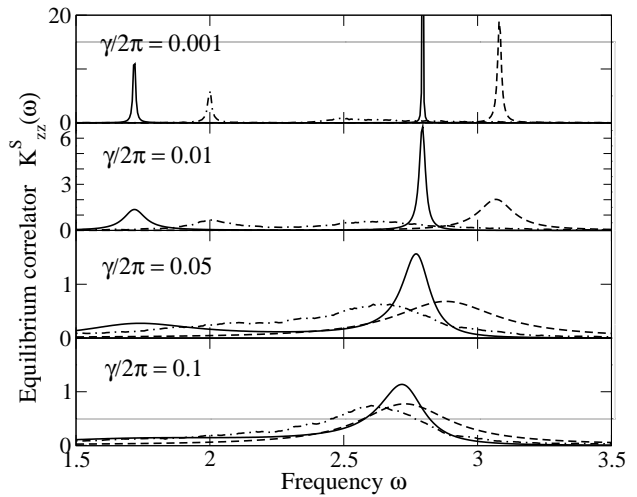}} \par}

\caption{\label{fig:gTplot}The Fourier-transform \protect\( K^{S}_{zz}(\omega )\protect \)
of the equilibrium correlator of \protect\( \hat{\sigma }^{S}_{z}(t)\protect \),
for different values of the noise strength $\gamma/(2\pi)$ (=0.001,
0.01, 0.05 and 0.1; from topmost to lowest graph). The values of the other
parameters are: \protect\( \Delta =1,\, \Delta _{S}=1.2, J =0.5 \)
and \protect\( \epsilon _{S}=0\protect \). Approach 1 and 4: solid line, approach 2: dashed line, approach 3: dash-dotted line. }
\end{figure}

The relevant parameters in our model are \( \epsilon _{S},\, \Delta _{S},\, \Delta, \) the coupling strength \( J \),
and the strength \( \gamma \) of the noise force $F$.
We choose the time scale such that \( \Delta \equiv 1 \).
The results discussed in the following have been calculated for \( \Delta _{S}=1.2 \) ($S$ and $B$ ``almost in resonance''). 

% generic features of mods. 1&2: number of peaks, Lorentzians
To begin our discussion, we note some generic features of the results obtained for approaches 1 and 2. Since in these cases $K_{zz}^S(\omega)$ is essentially the Fourier transform of a density matrix relaxing according to a master equation, it consists of several Lorentzian peaks. Their number is constrained to be less than the maximum number of transition frequencies of the respective system ($6$ for $S+B$ and $1$ for $S$ alone, plus possible zero-frequency ``pure'' relaxation). In practice, degeneracies between transition frequencies and selection rules reduce that number to $2$ (or $3$) for approach~1, and $1$ (or $2$) for approach~2, for $\epsilon_S=0$ (or $\epsilon_S\neq 0$). 

\begin{figure}
{\par\centering \resizebox*{0,87\columnwidth}{!}{\includegraphics{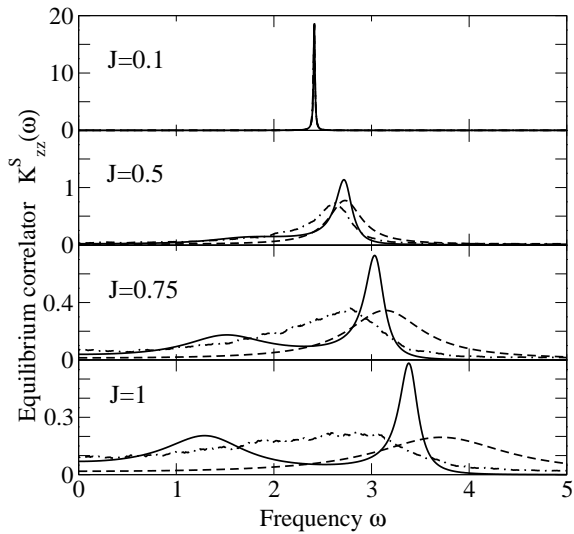}} \par}

\caption{\label{fig:eps0}The Fourier-transform \protect\( K^{S}_{zz}(\omega )\protect \)
of the equilibrium correlator of \protect\( \hat{\sigma }^{S}_{z}(t)\protect \),
for different values of the coupling strength \protect\( J\protect \) (=0.1,
0.5, 0.75 and 1.0; from topmost to lowest curve). The values of the other
parameters are: \protect\( \Delta =1,\, \Delta _{S}=1.2,\, \frac{\gamma}{2\pi}=0.1\)
and \protect\( \epsilon _{S}=0\protect \). Approach 1 and 4: solid line, approach 2: dashed line, approach 3: dash-dotted line. }
\end{figure}

\begin{figure}
{\par\centering \resizebox*{0,87\columnwidth}{!}{\includegraphics{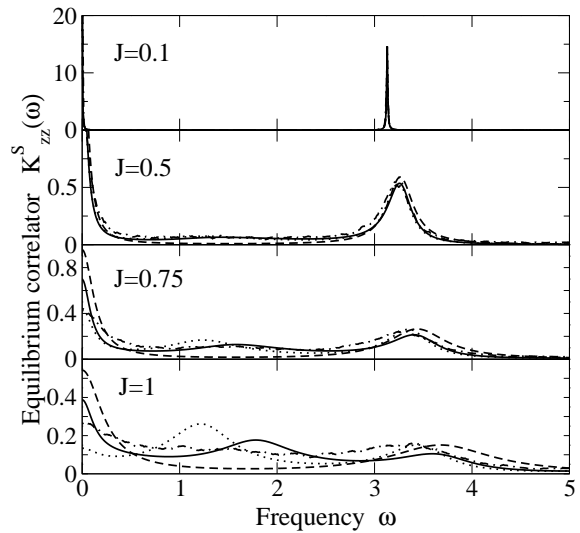}} \par}

\caption{\label{fig:eps1}The Fourier-transform \protect\( K^{S}_{zz}(\omega )\protect \)
of the equilibrium correlator of \protect\( \hat{\sigma }^{S}_{z}(t)\protect \),
for different values of the coupling strength \protect\( J\protect \) (=0.1,
0.5, 0.75 and 1.0; from topmost to lowest curve). The values of the other
parameters are: \protect\( \Delta =1,\, \Delta _{S}=1.2,\, \frac{\gamma}{2\pi}=0.1\protect \)
and \protect\( \epsilon _{S}=1\protect \). Approach 1: solid line, approach 2: dashed line, approach 3: dash-dotted line, approach 4: dotted line.}
\end{figure}

In the limit of weak coupling, $J\rightarrow 0$, all that remains is a
broadened peak at the transition frequency $2\Delta_S$ of system $S$
alone. In that limit, the results for all three models coincide, as
expected (see Figs. \ref{fig:eps0} and \ref{fig:eps1}). With
increasing $J$, the peaks get broadened and shifted, and additional
peaks may appear (in the case of approaches 1,3 and 4). 

% dependence on J: shift, width, size, coincidence
Naturally, the behaviour of approach 2 is simplest to analyze, since
it is the textbook example of a master equation applied to a single
two-level system. Since the correlator $\left\langle BB
\right\rangle_{\omega}$ is proportional to $J^2$, both the shift of
the transition frequency and the width of the peak(s) increase like
$J^2$, for arbitrarily large $J$. In contrast, the dependence of the peak
width and the frequency shift on the noise strength $\gamma$ is
non-monotonous. It is determined by the evolution of $\left\langle BB
\right\rangle_{\omega}$ (see Eq. (\ref{BB}) and Fig. \ref{fig:bbcorr})
with increasing $\gamma$. For very small $\gamma$, the two-level
fluctuator $B$ performs very weakly damped oscillations at the
frequency $2\Delta$. Unless it is exactly at resonance with system
$S$, the dissipative effects of $B$ on the dynamics of $S$ will be
weak in that regime. The decay rate of $S$, which is given by the
power spectrum of $B$ evaluated at $2\Delta_S$,  grows linearly in
$\gamma$ (for $\gamma^2\ll(\Delta_S^2-\Delta^2)^2/\Delta_S^2$). The
transition frequency of $S$ is shifted upwards or downwards,
depending on whether the main weight of the spectrum of \(B\) is
located below or above \(\Delta_{S}\)$(\Delta<\Delta_S$ or
$\Delta>\Delta_S)$. For increasing $\gamma$, $B$ performs more
strongly damped oscillations. The spectrum $\left\langle BB
\right\rangle_{\omega}$ concentrates around zero frequency (see
Fig. \ref{fig:bbcorr}) such that the decay rate of $S$ {\em decreases} again (like $1/\gamma$), after having gone through a maximum. The magnitude of the energy shift will also decrease for increasing $\gamma$, simply because the contributions of the power spectrum of $B$ lying to either side of $2\Delta_S$ will tend to cancel each other. However, in the limit $\gamma\rightarrow\infty$, the shift always saturates at a positive value which is independent of $\Delta$. These facts can be read off from the analytical result for approach 2 (written down in the special case of $\epsilon_S=0$):

\begin{equation}
K_{zz}^{S}(\omega )=\frac{1}{2\pi}\sum_{s=\pm 1} \frac{\Gamma}{\Gamma^2+(\omega-s \omega_0)^2},
\end{equation}

Here the peak width is given by \( \Gamma=2 J^2 \Re e \Sigma(2 \Delta _{S})=2 \pi\left\langle BB \right\rangle_{2\Delta_{S}} \)
%=(J^2\Delta^2\gamma)/((\Delta_{S}^2-\Delta^2)^2+\Delta_{S}^2\gamma^2) \)
, the shifted transition frequency is \( \omega_0=2 \Delta _{S}-2 J^2 \Im m  \Sigma(2 \Delta _{S}) \), and we have defined \( \Sigma(\omega)\equiv(2\gamma+i\omega)/(-\omega^2+4\Delta^2+2i\omega\gamma)\). 

% validity of master eq. 
The simple master equation (approach 2) is expected to come close to
the true result as long as the conditions of the Markoff and secular 
approximation are fulfilled. This means the coupling strength $J$ has
to be so small that the resulting decay of $S$ proceeds slowly compared
with the transition frequency itself (secular approximation) and
with the correlation time of the bath (Markoff approximation). The latter is given by  \(\tau_{B}=\frac{1}{\gamma} \) if \( \gamma^2 < 4 \Delta^2\) and \( \tau_{B}=\frac{1}{\gamma - \sqrt{\gamma^2-4 \Delta^2}}\) if  \( \gamma^2 > 4 \Delta^2\).

Now we pass on to a discussion of approach~1, which constitutes the  
exact solution for our physical system. 
The most notable difference to approach~2 is the
appearance of a second peak at the transition frequency $2\Delta$ of the
two-level fluctuator $B$. At small $J$, the strength of this peak grows
like $J^2$, while its width is fixed (depending on $\gamma$). In this way,
the power spectrum of the bath fluctuations shows up in the short-time
behaviour of the correlator of the system $S$. This behaviour cannot 
be captured by the
master equation (approach 2). Increasing $J$ leads to a frequency shift and 
a change in the width of the ``original'' peak at $2\Delta_S$, much like
predicted by the simpler approach 2. However, in the description of approach~1,
 these changes are due to the change in eigenfrequencies and eigenvectors
of the combined system $S+B$. At small $J$, the results of approaches 1 and 2
can be shown to coincide using perturbation theory (compare also
Fig. \ref{fig:eps0} and \ref{fig:eps1}, topmost graphs). Deviations from approach~2 appear at higher
values of $J$, where the energy shift of approach~1 only grows {\em
linearly} with $J$ (see Figs. \ref{fig:eps0} and \ref{fig:eps1}, lower graphs).
In contrast, the frequency of the second peak is suppressed to zero. This
behaviour can easily be found from the diagonalization of the Hamiltonian
for the combined system $B+S$ in the limit $J\rightarrow \infty$, when
one obtains two pairs of degenerate energy levels, separated by $2J$.

Regarding the dependence on the noise strength $\gamma$, the same
qualitative remarks apply as for approach~2. However, it is interesting
to note that there {\em is} a frequency  shift with increasing $\gamma$ in
approach 1 as well (see Fig. \ref{fig:gTplot}), in spite of the fact that the additional terms in
the nonsecular master equation (\ref{mastereqnonsec}) seem to describe
a purely relaxational dynamics. This is in contrast to the behaviour
known from the usual form of the master equation, Eq. (\ref{mastereq}),
where the energy shifts can be read off directly from the {\em imaginary}
coefficients in the equation.

\begin{figure}
{\par\centering \resizebox*{0,95\columnwidth}{!}{\includegraphics{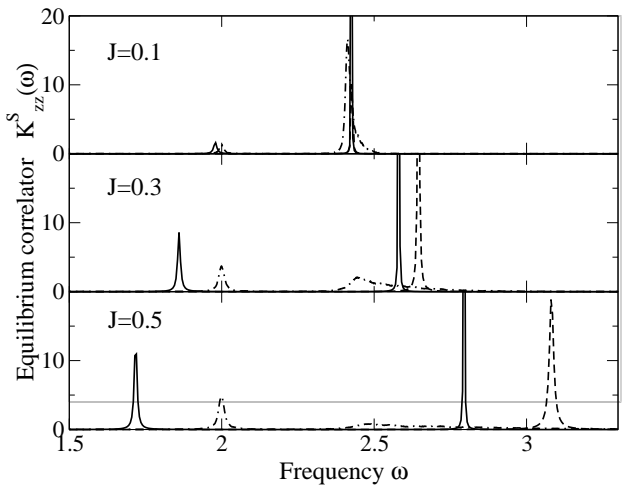}} \par}

\caption{\label{fig:smallJ}The Fourier-transform \protect\( K^{S}_{zz}(\omega )\protect \)
of the equilibrium correlator of \protect\( \hat{\sigma }^{S}_{z}(t)\protect \),
 for different values of the coupling strength \protect\( J\protect \) (=0.1,
0.3 and 0.5; from topmost to lowest curve). The values of the other
parameters are: \protect\( \Delta =1,\, \Delta _{S}=1.2,\, \frac{\gamma}{2\pi}=0.001\protect \)
and \protect\( \epsilon _{S}=0\protect \). Approach 1 and 4: solid line, approach 2: dashed line, approach 3: dash-dotted line.}
\end{figure}

% lin. vs. nonlin.: asymm. peaks, no shifts in lin.bath model
We now turn to approach~3, where the nonlinear bath has been replaced
by a linear bath (i.e. a colored Gaussian random process in our case).
Since this model takes the full bath spectrum \(\left\langle BB \right\rangle_{\omega }\) as input, this spectrum may also show up in the
result for the system correlator $K_{zz}^S(\omega)$, as is indeed
the case. Fig. \ref{fig:gTplot} demonstrates that this effect 
is most pronounced for small values of $\gamma$, where
the bath spectrum has a relatively sharp structure (the noise field
$B(\cdot)$ acting on $S$ deviates strongly from white noise). In these
cases, the qualitative agreement between approach~3 and approach~1 (``exact
solution'') is much better than that between approach~2 (``simple
master equation'') and approach 1 (see also Fig. \ref{fig:smallJ}). Nevertheless, there are deviations: In particular,
there is no visible shift of the peaks in approach~3 with increasing $J$.
They just become wider and asymmetric (this applies especially to the 
peak at frequency $2\Delta_S$). For higher values of $\gamma$,
the linear bath (approach~3) in general shows less structure than the
exact solution, obtained for the actual nonlinear bath. 

Finally, we discuss approach 4, where the second peak shows up, in contrast to the Markoff
approximation. In general, we would expect the weak-coupling solution to be a bit worse
than the simulation of the linear bath with colored noise
correlations (approach 3), since it is an approximation to the latter
case. However, the solution for the special case
\(\epsilon_{S}=0\) turns out to coincide completely with the exact
solution (approach 1).
The solution for \(\epsilon_{S}=1\) (or, more generally,  \(\epsilon_{S}\ne 0\)) is
good for small system-bath coupling \(J\). 
It fails for increasing \(J\), where approach 3 seems to be the better
approximation, provided \(\epsilon_{S}\) is not too small (see
discussion above).

\section{Conclusions}

We have discussed a simple model of a nonlinear bath, consisting of a single two-level
system subject to a classical white-noise force. Its action on another two-level
system has been analyzed using four different approaches. 
The exact evolution of the density matrix has been compared with the
results obtained using a standard master equation, a linear bath (colored noise) which has been substituted for the
nonlinear bath, and a weak-coupling equation.
Numerical results for various special cases have been discussed. 
For strong system-bath coupling, the linear bath may still be a good
approximation in a regime where the simple master equation already fails.
This applies in particular when the bath spectrum has a strongly peaked
structure. However, deviations between
the linear and the original nonlinear bath are clearly visible. 

It is hoped that, in a future work, also the (technically more involved)
case of arbitrary finite temperature may be discussed. Further
possible extensions include an analysis of the higher-order terms in
the weak-coupling equation as well as replacing \(B\) by a spin of
larger magnitude, in order to observe the transition to the linear
bath.

\begin{acknowledgments}

We like to thank H.-A. Engel, G.-L. Ingold, D. Loss,  P. Talkner,
W. Zwerger,  and particularly F. Meier for helpful discussions. Our work was supported by the Swiss NSF and the
NCCR Nanoscience.
\end{acknowledgments}

\end{document}